\documentclass{iau}
\usepackage{graphicx}

\title[Accretion disk parameters in HLX-1] 
{Accretion disk parameters in HLX-1}

\author[Roberto Soria \& George Hau]   
{Roberto Soria $^1$
 \and  George Hau $^2$}

\affiliation{$^1$ International Centre for Radio Astronomy Research, 
Curtin University, \\
GPO Box U1987, Perth, WA 6845, Australia \\ 
email: {\tt roberto.soria@icrar.org } \\[\affilskip]
$^2$European Southern Observatory, Alonso de Cordova 3107, Santiago, 
Chile \\email: {\tt ghau@eso.org}}

\pubyear{2012}
\volume{290}  
\jname{Feeding compact objects: Accretion on all scales}
\editors{C.M. Zhang, T. Belloni, M. M\'endez \& S.N. Zhang, eds.}

\begin{document}

\maketitle

\begin{abstract}
We estimate the outer radius of the accretion disk in HLX-1 from its optical 
brightness and from the exponential timescale of the decline 
in the X-ray lightcurve after an outburst. We find that the disk is 
an order of magnitude smaller than the semimajor axis of the orbit.
If the disk size is determined by the circularization radius near periastron, 
the eccentricity of the binary system must be $\gtrsim 0.95$.
We report on the discovery of H$\alpha$ emission during the 2012 outburst, 
with a single-peaked, narrow profile (consistent with a nearly face-on view),
and a central velocity displaced by $\approx 490$ km s$^{-1}$ 
from that of the host galaxy.

\keywords{accretion, accretion discs -- X-rays: individual: HLX-1 -- black hole physics}
\end{abstract}

\vspace{-0.2cm}

\firstsection 
\vspace{-0.2cm}
\section{Introduction}
\noindent
The X-ray source 2XMM\,J011028.1$-$460421 
(HLX-1) is the strongest intermediate-mass black hole 
(IMBH) candidate known to date (\cite{far09}).
It is located at a projected distance of $8'' \approx 3.7$ kpc 
from the nucleus of the S0 galaxy ESO\,243-49 ($z = 0.0224$, 
$d \approx 95$ Mpc).
If the peak X-ray luminosity $\approx 10^{42}$ erg s$^{-1}$ 
is Eddington limited, the BH mass has to be $\sim 10^4 M_{\odot}$. 
A similar value is obtained by modelling the X-ray spectrum 
with thermal disc emission (\cite{far09,dav11,ser11}).
Its X-ray spectral variability (\cite{god09,ser11}) 
and its radio flares detected in association with the X-ray 
outbursts (\cite{web12}) are consistent with the canonical 
state transitions and jet properties of an accreting BH. 
HLX-1 has a point-like, blue optical counterpart 
($B \sim V \sim 24$ mag near the outburst peak; \cite{far12,sor12,sor10}).
The H$\alpha$ emission line at a redshift consistent with that 
of ESO\,243-49 (\cite{wie10}) is the strongest argument for 
a true physical association. It is still debated whether the optical/UV 
continuum is dominated by the irradiated accretion 
disc, or by a young star cluster around the BH (\cite{sor12,far12,map13}).
The X-ray flux shows recurrent outbursts every $\approx 370$ d, 
probably triggered by a sudden increase 
in the mass trasnfer rate when the donor star is near periastron 
of a highly eccentric orbit (\cite{las11}). In this scenario, 
we can use the X-ray lightcurve to constrain the system parameters.



\begin{figure}[t]
\begin{center}
 \includegraphics[width=2.6in,angle=270]{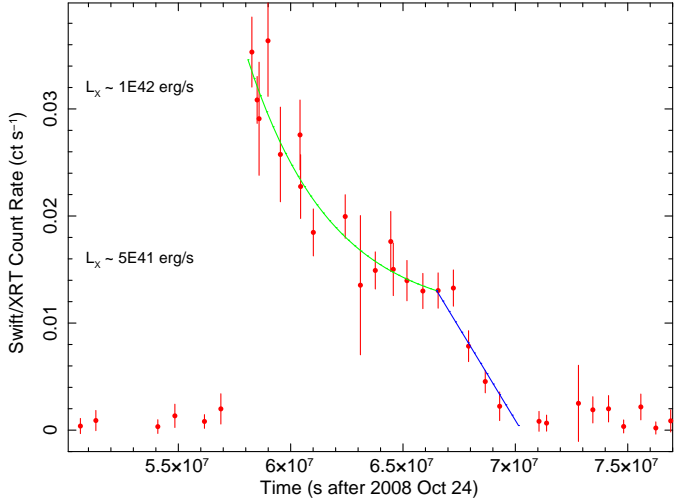}
 \caption{{\it Swift}/XRT lightcurve of the 2010 outburst, fitted with 
a standard X-ray transient model (exponential decay, knee, linear decay).
It is remarkably similar to those of several transient Galactic 
X-ray binaries modelled by \cite[Powell et al. (2007)]{pow07}.}
   \label{fig1}
\end{center}
\end{figure}

\vspace{-0.4cm}

\section{Disc size}
\noindent
Fits to the {\it XMM-Newton}, {\it Chandra}, {\it Swift} spectra 
during outbursts have consistently shown that 
the size of the inner disk (with 
$kT_{\rm in} \approx 0.2$ keV)
is $\sim$ a few $10^9$ cm (\cite[Farrell et al. 2009]{far09},
\cite{ser11,dav11,sor11,sor12},\cite[Farrell et al. 2012]{far12}), 
consistent with an IMBH.
Instead, much less is known about the outer disc radius $R_{\rm out}$.
If the disc is the dominant optical emitter, the {\it HST} 
and VLT studies 
of \cite[Farrell et al. (2012)]{far12} 
and \cite[Soria et al. (2012)]{sor12} agree on 
$R_{\rm out} \lesssim 2 \times 10^{13}$ cm for a face-on disc. 
A recent reanalysis of the {\it HST} data (\cite{map13}) suggested  
$R_{\rm out} \approx 3.5 \times 10^{13}$ cm for 
a viewing angle $i=45^{\circ}$ (a little smaller if face-on).



There is an alternative way to estimate the outer disc size, based 
on the X-ray outburst decline timescale. 
Following \cite[King \& Ritter (1998)]{kin98} and 
\cite[Frank et al. (2002)]{fra02}, we assume that the outbursting disc 
is approximately in a steady state with surface density
$\Sigma \equiv \rho H \approx \dot{M}_{\rm BH}/(3\pi \nu)$, 
where $\dot{M}_{\rm BH}$ is the central accretion rate and $\nu$ 
the kinematic viscosity.
When the whole disc from $R_{\rm in}$ to $R_{\rm out}$ is in a hot, 
high-viscosity state, its total mass 
\begin{equation}
M_{\rm disc} = 2\pi \int_0^{R_{\rm out}} \Sigma R\,dR 
   \approx \frac{\dot{M}_{\rm BH} R^{2}_{\rm out}}{3\nu} 
   = \frac{-\dot{M}_{\rm disc} R^{2}_{\rm out}}{3\nu},
\end{equation}
where $\nu$ is the kinematic viscosity near the outer edge 
of the disc (\cite{kin98}).
Integrating Eq.(2.1) gives an exponential decline 
for the disc mass, the accretion rate,
and the outburst luminosity $L \sim L_{\rm X} \sim 0.1\dot{M}_{\rm BH}c^2$. 
We expect to see a luminosity  
\begin{equation}
L_{\rm X} \approx L_{{\rm X},0} \exp(-3\nu t/R^{2}_{\rm out}),
\end{equation}  
where $L_{{\rm X},0}$ is the value at the outburst peak, declining 
on a timescale 
$\tau_{\rm e} \approx R^{2}_{\rm out}/(3\nu)$,
as long as the rate at which the disc mass is depleted during 
the outburst decline is much larger 
than any ongoing transfer of mass from the donor star.
For the viscosity, we take 
the usual parameterization $\nu = \alpha c_{\rm s} H$ (\cite{sha73}),
where $\alpha$ is the viscosity coefficient in the hot state, 
$c_{\rm s}$ is the sound speed, and $H$ the vertical scaleheight. 
In the Shakura-Sunyaev disc solution (\cite{sha73,fra02}), 
with Kramers opacity, after some algebra we obtain (\cite{sor13}):
\begin{equation}
R_{12} \approx \left(\frac{\tau_{\rm e}}{5.2 \times 10^6}\right)^{4/5} \, 
    \alpha^{16/25} \dot{M}^{6/25}_{22} m_3^{-1/5},
\end{equation}
where $\dot{M}_{22}$ is the accretion rate in units of $10^{22}$ g s$^{-1}$, 
$m_3$ is the BH mass in units of $10^3 M_{\odot}$, 
$R_{12} \equiv R_{\rm out}/(10^{12} \rm{cm})$.

The exponential decay continues until the outer disc annuli can no longer 
be kept in the hot state, so that hydrogen recombines and viscosity drops. 
From that moment, the central accretion rate declines linearly (\cite{kin98}), 
with a slope such that
$t_{\rm end} - t_1 = \tau_{\rm e}$,
where $t_{\rm end}$ is the (extrapolated) time in which 
the accretion rate and luminosity go to zero.
Finally, if there is ongoing mass 
transfer $\dot{M}_2$ from the donor star during the outburst, 
the lightcurve has a characteristic ``knee'' at the point 
where it switches from an exponential to a linear decline (\cite{pow07}).

\cite[Soria (2013)]{sor13} fitted the {\it Swift} X-ray lightcurve 
of the 2010 outburst (Fig.~1) to obtain two independent estimates 
of the viscous timescale $\tau_{\rm e}$, from the exponential and 
the linear regime, and used them to constrain $R_{\rm out}$.
For the exponential part, 
$\tau_{\rm e} = R^{2}_{\rm out}/(3\nu) = 3.7^{+5.0}_{-1.5} \times 10^6$ s 
($90\%$ confidence limit). 
For the linear part, $\tau_{\rm e} = 3.5^{+1.0}_{-0.8} \times 10^6$ s.
Assuming a peak accretion rate $\dot{M} \approx 2\times 10^{22}$ g s$^{-1}$ 
and a viscosity parameter $\alpha \lesssim 1$, and more likely 
$\alpha \sim 0.3$ (\cite{fra02}), both decline timescales give 
$R_{\rm out} \sim 10^{12}$ cm.
We also determined e-folding decline timescales $\approx 5 \times 10^6$ s and 
$\approx 3 \times 10^6$ s for the 2009 and 2011 outbursts, respectively, 
both consistent with an outer radius $\sim 10^{12}$ cm, 
in the standard disc approximation.
Those values are an order of magnitude smaller than what is estimated 
by assuming that most of the continuum emission comes from the hot disc; 
and the latter is already a small size compared with 
the semi-major axis.

\vspace{-0.4cm}

\section{Orbital parameters}
\noindent
We can now compare the size of the disc estimated from optical and X-ray 
flux measurements 
($10^{12}$ cm $\lesssim R_{\rm out} \lesssim 3 \times 10^{13}$ cm) with 
the characteristic size of the binary system. 
The semimajor axis $a$ of the binary is 
\begin{equation}
a = 1.50 \times 10^{13} m^{1/3} (1+q)^{1/3} P_{\rm yr}^{2/3} 
    \ {\rm {cm}},
\end{equation}
where $q = M_2/M_{\rm BH}$ 
and $m \equiv (M_{\rm BH} + M_2)/M_{\odot} \equiv M/M_{\odot}$. 
Typical values for HLX-1 in the IMBH scenario are 
$q \sim 10^{-3}$ and $m^{1/3} \sim 10$--$20$. Therefore, 
the semimajor axis is at least 5, 
and possibly up to 100 times larger than the disc radius. 
This mismatch clearly suggests an eccentric orbit (\cite{las11}), 
in which the characteristic disc size is determined by the periastron 
separation $R_{\rm per} = (1-e) a$ and the tidal truncation radius 
$R_{\rm out} \approx 0.6 (1-e) a$ (\cite[Warner 1995]{war95}). 
For $R_{\rm out} = 10^{13}$ cm, the tidal radius condition 
requires an eccentricity $e \approx 0.89$ for $M_{\rm BH} = 1000 M_{\odot}$, 
and $e \approx 0.95$ for $M_{\rm BH} = 10^4 M_{\odot}$.

We can argue that the tidal truncation constraint is not relevant 
here, if mass transfer occurs impulsively only near 
periastron: the disc may be small because it does not have 
time to grow to its tidal truncation radius before being accreted.
Instead, the circularization radius provides a stronger 
{\it lower limit} to the predicted disc size, and is applicable to any system 
where mass transfer occurs through the Lagrangian point L$_1$.
The circularization radius $R_{\rm cir}$ is defined via the conservation 
of angular momentum equation 
$v_{\phi}\left(R_{\rm cir}\right) R_{\rm cir} = \left(X_{\rm L1}R_{\rm per}\right)^2 
\Omega \left(R_{\rm per}\right)$,
where $v_{\phi}$ is the orbital velocity of the accretion stream around the BH, 
$X_{\rm L1}R_{\rm per}$ the distance between the BH and L$_1$, and 
$\Omega \left(R_{\rm per}\right)$ the angular velocity of the donor star 
at periastron.
We computed $X_{\rm L1}$ from Eq.(A13) of \cite[Sepinsky et al. (2007)]{sep07}, 
valid 
for eccentric orbits, and found (\cite{sor13}) $X_{\rm L1} \sim 0.95$ 
for the expected mass ratio and period.
For $M_{\rm BH} = 5 \times 10^3 M_{\odot}$, a circularization 
radius $R_{\rm cir} = 10^{13}$ cm requires $e \approx 0.97$.
This eccentricity is much more extreme than what was suggested in 
\cite[Lasota et al. (2011)]{las11}.
It may seem implausible, knowing that tidal forces tend 
to circularize orbits in X-ray binaries.
However, \cite[Sepinsky et al. (2007,2009)]{sep07,sep09} showed that 
in the case of a donor star that transfers mass impulsively only at periastron, 
with $q \lesssim 1 -0.4e +0.18e^2$, the secular evolution of the orbit leads 
to an increase of both eccentricity 
and semimajor axis, even when the opposite effect 
of tidal forces is taken into account.

We also assessed whether the donor stars can avoid tidal disruption 
for such a small periastron distance. Comparing this distance to 
the tidal disruption radius (\cite{ree88}), we find that the star survives if
$M_2 \gtrsim 4.6 \times 10^{-5} M_{\rm BH}$, 
easily satisfied in the likely mass range of HLX-1.
The small periastron distance sets an upper limit to the radius 
of the donor star: $R_2 \lesssim$ a few $R_{\odot}$, ruling out supergiants, 
red giants and AGB stars. Possible donors are main sequence 
(B type or later) or subgiants. 


There is at least one class of stellar objects where eccentricities 
$\gtrsim 0.95$ are common:
S stars observed on highly eccentric orbits within 0.01 pc of the Galactic 
nuclear BH (\cite{gil09}). A possible scenario for the origin 
of Galactic S stars is the tidal disruption of a stellar binary system 
near the BH, which produces an escaping, hyper-velocity star, and 
a more tightly bound star on a very eccentric orbit.

\vspace{-0.4cm}

\section{H$\alpha$ emission line}

Galactic BHs in outburst usually show optical emission lines 
({\it eg}, H$\alpha$, H$\beta$, He{\footnotesize{II}} 4686) 
from the irradiated outer disk, often with a double-peaked profile. 
The peak separation or 
full-width-half-maximum (FWHM) provides an estimate of the projected 
velocity of the outer disk (where such lines are mostly emitted), 
and therefore constrains $R_{\rm out}$.
We observed HLX-1 with VLT/FORS on 2012 Aug 27--28, and Sep 11,
near outburst peak. We will present a full discussion 
of the results in a paper currently in preparation.

Our main preliminary result is that we found a strongly significant 
H$\alpha$ emission line on all three nights, strengthening the findings  
of \cite[Wiersema et al. (2010)]{wie10}.
The central wavelength of the line (averaged over all 3 nights) 
is $\lambda \approx 6720.45$ \AA, that is a velocity of 7205 km s$^{-1}$, 
redshifted by $\approx 490$ km s$^{-1}$ with respect to the central 
velocity of ESO243-49. For an S0 galaxy of similar size, we expect 
a maximum stellar rotational velocity $\sim 250$--$350$ km s$^{-1}$, 
from the Tully-Fisher relation (\cite{bla09}). 
If HLX1 is the remnant of an accreted satellite dwarf 
(\cite{map12}), it must be moving on an eccentric orbit.
The H$\alpha$ line profile is remarkably narrow and single peaked, 
with a Gaussian FWHM $\approx 10.5$ \AA\ $\approx 480$ km s$^{-1}$. 
The rotational velocity of a Keplerian orbit around the BH 
at $R = 10^{13}$ cm 
is $v_{\phi} \approx 1140 \, m_3^{1/2}$ km s$^{-1}$; 
therefore, a disk line should have 
a FWHM $\approx (2300 \, m_3^{1/2} \sin i)$ km s$^{-1}$. 
This suggests that either the disk is seen face on 
($i \lesssim 10^{\circ}$), or the line is not from 
a rotating thin disk but from an extended envelope.

\end{document}